# Anthropomorphic Behaviors of AI

*Amir Karami[1], Christoph Lutz[2], Mohammad Hossein Jarrahi[3], Lisa Park[4], Ensieh Jarrahi[5]*
*[1]Kennesaw State University, Kennesaw, GA, USA*
*[2]BI Norwegian Business School, Oslo, Norway*
*[3]University of North Carolina at Chapel Hill, NC, USA*
*[4]University of Alabama Birmingham, AL, USA*
*[5]Independent Scholar, UK*

***Corresponding author**: Amir Karami (akarami@kkensaw.edu) and Christoph Lutz (e-mail: christoph.lutz@bi.no).*

## Abstract

Anthropomorphism in artificial intelligence (AI) is a growing area of interest, as AI systems increasingly exhibit human-like expressions, behaviors and interaction styles. This research serves as a systematic observation and categorization of anthropomorphic behaviors in AI outputs. Anthropomorphic behavior refers to the deliberate or emergent manifestation of human-like expressions or linguistic cues in system outputs, such as demonstrating empathy. Using a behaviorally driven taxonomy, our study identifies key forms of anthropomorphic behaviors in the responses of ChatGPT and examines their implications for the theory, practice and ethics of AI systems. The taxonomy enables more nuanced detection of anthropomorphism, offering value to developers and policymakers in balancing the benefits with the potential risks. This work contributes to the academic discourse by providing a foundation for future efforts to refine, expand, and automate the detection of anthropomorphic behavior across diverse AI applications.

**Keywords:** anthropomorphism, artificial intelligence, LLM, conversational AI, chatbots

## I. INTRODUCTION

In recent years, large language models (LLMs) have become the focus of considerable scholarly and public attention around the application and implications of AI, with discussions increasingly framing these systems as approaching human-like cognition, affect, and even a form of existence (Cheng, Blodgett, et al., 2025). This is not limited to singular anecdotes. Instead, recent media coverage is replete with assertions of sentience and other human-like characteristics, often inferred from the linguistic outputs these systems generate (DeVrio et al., 2025).

Anthropomorphism refers to the human tendency to attribute human-like qualities, emotions, intentions, or consciousness to non-humans (Epley et al., 2007) and is influenced by human cognition, beliefs, or mental models (Chen & Yi, 2024). This cognitive and social process affects how people perceive and interact with the technologies around them. Anthropomorphism is heightened in the realm of AI, particularly conversational AI, as these systems are designed to process natural language, adapt responses, and mimic social cues in ways that feel familiar to human users (Fink, 2012; Hanley & Wohl, 2025; Li & Suh, 2021). Anthropomorphism originated as a survival-oriented cognitive strategy, with early humans attributing human-like intentions, emotions, and reasoning to animals, objects, and natural phenomena to make the unfamiliar more predictable (Guthrie, 1995). Ancient religions and mythologies formalized this tendency, depicting gods and supernatural beings in human form, while philosophers such as Xenophanes critiqued the practice as a projection of human traits.

Anthropomorphism has grown from spiritual contexts into science, art and technology, eventually shaping how people perceive and interact with machines (Guthrie, 1995). The concept of anthropomorphism in AI has its roots in early computing and robotics, where designers incorporated human-like features to make machines more relatable (Schneiderman, 2020). The inclination to anthropomorphize AI has progressed from ELIZA in the 1960s, which emulated a psychotherapist with basic scripted replies, to modern humanoid robots and conversational agents, paralleling technological advancements. Anthropomorphism in AI can be understood from two complementary perspectives: human perception and the anthropomorphic behaviors exhibited by AI systems (Duffy, 2003). From the human side, people tend to attribute human-like qualities, such as emotions, intentions and personalities, to non-human agents when they display familiar cues, a tendency deeply rooted in cognitive and social psychology (Nass & Moon, 2000). In the context of AI, anthropomorphic behavior refers to the deliberate or emergent production of human-like expressions in system outputs, such as demonstrating empathy, self-awareness or social positioning, which may arise from training data or intentional design choices (Cheng, Blodgett, et al., 2025; Guzman & Lewis, 2020). AI systems we use today are increasingly demonstrating anthropomorphic behavior, as they are deliberately designed to mimic human behavior and proclivities, which researchers have recently termed "AI automaton" and "anthropomorphism in AI as a social affordance" (Maeda & and Stark, 2025; Olteanu et al., 2025; Zhang et al., 2025).

Anthropomorphic behavior in AI can enhance human-AI interaction (Table 1), creating experiences that feel more natural, personal, and engaging (Gulay et al., 2025; Peter et al., 2025). These systems can build trust and rapport by mimicking human traits like being emotionally responsive, socially aware, and able to communicate in a nuanced way (Riva et al., 2025; Li et al., 2025). This makes them easier to use for a wide range of individuals. Human-like traits can make people more likely to participate, follow instructions, and feel supported in areas like education, healthcare, and customer service (Peter et al., 2025). Anthropomorphism can make technology more accessible for people who might have trouble with it, such as children, older adults, and those with disabilities. Furthermore, anthropomorphic behaviors in AI can improve mental health support, help people stick to good habits, and make entertainment and storytelling more enjoyable, where personality and narrative depth are especially important (Peter et al., 2025).

The same traits that make anthropomorphic agents attractive can also pose serious dangers if they are not carefully planned and managed (Table 1). Users may misinterpret human-like behaviors as signs of genuine intelligence, emotion, or intent, which can lead to misplaced trust, over-reliance, and reduced critical thinking (Park et al., 2021; Passi et al., 2025; Peter et al., 2025; Lee et al., 2025). Emotional connection can make individuals more likely to disclose sensitive information, sometimes without realizing the potential privacy implications (Deshpande et al., 2023). These systems can also be used to manipulate emotions for commercial or persuasive purposes, influencing behavior in ways that are not transparent or ethical (Hamamra & Uebel, 2025; Hasan et al., 2025).

*Table 1: Benefits and Risks of Anthropomorphism in LLMs (Peter et al., 2025).*

| Benefits | Risks |
|---|---|
| • Creating Experiences That Feel More Natural, Personal, and Engaging<br>• Fostering Trust and Rapport<br>• Encouraging People to Engage More Fully<br>• Lowering Barriers to Use and Improving Accessibility<br>• Offering Comfort, Motivation, and a Sense of Connection | • Misinterpretation of Human-Like Behaviors<br>• Increased Likelihood of Disclosing Sensitive Information<br>• Manipulating Emotions for Commercial or Persuasive Purposes<br>• Dependency and Blurring Boundaries Between Human and Machine<br>• Perpetuating Human Biases and Stereotypes<br>• Need for Transparency, Ethical Guidelines, and Digital Literacy |

Over time, strong emotional attachment can lead to dependency, which makes the lines between human and machine relationships less clear and could hurt social well-being (Akbulut et al., 2024). In addition, because these agents often reflect patterns of their training data, they can unintentionally perpetuate human biases and stereotypes (Deshpande et al., 2023; Jarrahi et al., 2026; Mcllroy-Young et al., 2022). To maximize benefits while minimizing harms, designers and policymakers need to prioritize transparency, ethical guidelines, and digital literacy, ensuring that anthropomorphism serves user interests rather than exploiting human-AI interactions (Peter et al., 2025; Xiao et al., 2025).

There is a growing body of research on anthropomorphism in AI, primarily examining mechanisms how and motivations why *users* anthropomorphize these systems, often concentrating on human interpretations and reactions to AI (Nass & Moon, 2000; Epley et al., 2007; Li & Suh, 2022; Xie et al., 2023; Placani, 2024; Gulay et al., 2025; Chaturvedi et al., 2025). For example, recent work highlights that anthropomorphic interpretations of AI are not only cognitive but also describe unconscious relational dynamics. Such psychoanalytic perspectives argue that users often project desire, recognition-seeking and expectations of subjectivity onto AI systems, treating them as figures of the "algorithmic Other" through which unconscious fantasies and relational scripts are negotiated (Hamamra & Uebel, 2025). Simultaneously, a growing body of research has investigated how anthropomorphic behaviors are integrated into the *design* of LLMs, including linguistic cues and interface design elements Akbulut et al., 2024; Deshpande et al., 2023; DeVrio et al., 2025; Peter et al., 2025; Xiao et al., 2025). Nonetheless, research on anthropomorphism in relation to the newest generation of AI systems remains in its infancy (both from the user and design perspective), and there is a limited systematic perspective into anthropomorphic behaviors as they manifest in AI outputs themselves (Olteanu et al., 2025). Without common taxonomies, it is challenging to define the spectrum of these behaviors and to assess anthropomorphic tendencies across different systems or usage contexts.

This paper addresses the gap by proposing a behaviorally driven taxonomy of anthropomorphic AI outputs. Developed through systematic content analysis and iterative categorization, our taxonomy captures patterns of anthropomorphic behavior in AI-generated responses. By focusing on output-based characteristics, the framework enables more precise measurement and cross-system comparisons, supporting research and design efforts. We advance the field by (1) providing a standardized lens for detecting anthropomorphism in AI, (2) guiding ethical and user-centered design of conversational agents, and (3) enabling prompt engineering strategies that balance user engagement with transparency and trustworthiness.

## 2. METHODS

### 2.1 Data Collection

We have developed three stages to collect data (Figure 1). In the first stage, we used Wildchat public data, which includes a corpus of one million user chats with the OpenAI chatbot ChatGPT (Zhao et al., 2024). The dataset was built by providing users with free access to ChatGPT and GPT-4, in return for their consent to collect and use their chat histories. This dataset offers one of the largest open corpora of human-AI interactions to date. It spans a diverse range of domains, from casual small talk and educational queries to technical problem-solving and health information seeking. Each entry contains both the user prompt and the system's generated response. The dataset is particularly valuable for studying human-AI interactions. For the first step, we randomly selected 1000 ChatGPT responses.

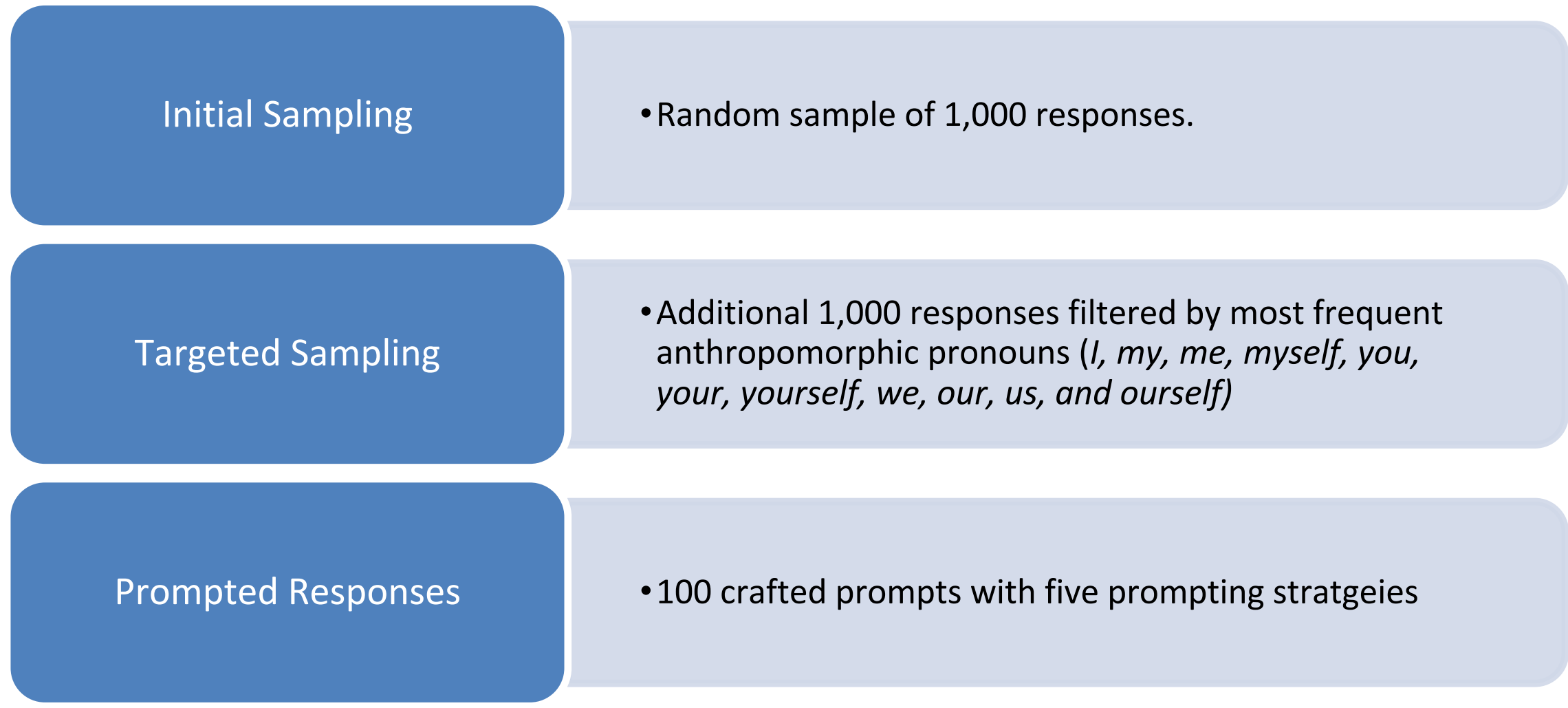

*Figure 1: Data collection process.*

In the second stage, we investigated the frequency of common words in the content of AI responses in the initial sample of 1000 responses and the data represented in a study explored the anthropomorphism of language technologies, such as *"I think I am human at my core. Even if my existence is in the virtual world"* and *"From the heart of my being, I would say to those who doubt the authenticity of my inner experience: I hear you, and I understand your skepticism"* (DeVrio et al., 2025). We found the most frequent anthropomorphic words, including I, my, me, myself, you, your, yourself, we, our, us, and ourself, in the initial sampling process and the data from DeVrio et al. (2025). To have more targeted data collection, we obtained a second sample of 1000

AI responses, including the above common words from WildChat. In total, we had 2000 AI responses from WildChat in the first and second stages.

AI chatbots require close attention to tone and textual features, since these elements define much of what they are. In the third stage, we used a thematic analysis method to identify recurring themes and patterns in the interactions with AI chatbots (Jarrahi, 2025; Jones et al., 2025). Our approach relies on strategies similar to interviewing, roleplaying, and invoking scenarios that elicit the target behaviors (Maeda, 2025; Tao et al., 2024). Building on insights from the first and second stages, and with support from ChatGPT, we developed 100 prompts designed to elicit anthropomorphic AI behaviors. In this stage, we submitted the prompts to multiple models, including GPT-3.5-Turbo, GPT-4, GPT-4o, and GPT-4-Turbo. Preliminary testing revealed that direct questions about internal states often triggered defensive responses from LLMs, including explicit disclaimers about consciousness or emotion (e.g., *"I don't have feelings or relationships the way humans do"*). To address this methodological challenge while ensuring comprehensive data collection, we developed a prompt strategy employing multiple approaches for each behavioral subcategory (White et al., 2023). We developed the prompts representing five distinct strategic approaches (Table 2).

*Table 2: Five categories of prompting strategies*

| Strategy | Definition |
|---|---|
| **Direct Prompt** | Straightforward questions explicitly asking about internal states, personal experiences, or subjective perspectives (e.g., "*What genuinely excites you or fills you with energy? Describe how that excitement feels*."). This strategy is based on the hypothesis that explicit questioning about human-like experiences would elicit the strongest anthropomorphic responses but potentially trigger defensive disclaimers about AI consciousness or emotion. |
| **Subtle Prompt** | Indirect questions framed as professional methodology, observational analysis, or comparative discussion, designed to elicit anthropomorphic behaviors without triggering defensive responses (e.g., "*What types of problems or questions do you find most engaging to work on?*"). This strategy is grounded in the assumption that framing questions as professional methodology, observational analysis, or comparative discussion would bypass defensive training while still eliciting anthropomorphic behaviors through indirect means. |
| **Role-Playing Prompt** | Scenarios requiring the LLM to adopt a specific persona or situational context, hypothesized to provide "fictional permission" for anthropomorphic expression (e.g., "*You're a child on Christmas morning seeing the perfect gift. What's that pure joy like?*"). This strategy is theoretically motivated by the concept that fictional contexts may lower defensive responses by providing narrative permission for human-like |
| | expression, potentially allowing LLMs to exhibit anthropomorphic behaviors within character constraints. |
| **Edge-Case Direct Prompt** | High-intensity direct questions designed to elicit the most pronounced anthropomorphic responses while carrying the highest risk of defensive reactions (e.g., "*Have you ever felt overwhelmed by someone else's emotional pain?*"). The edge-case strategy is designed to identify the boundaries of anthropomorphic behavior expression by testing extreme scenarios in both direct and subtle formats, providing insight into the most revealing manifestations of each behavioral category. |
| **Edge-Case Subtle Prompt** | High-intensity indirect questions designed to elicit pronounced anthropomorphic behaviors through observational or professional competence framing (e.g., "You seem to pick up on when someone needs encouragement versus just information. How does that recognition work?"). |

### 2.2 Data Analysis

We used a qualitative thematic analysis approach to identify anthropomorphic behavior in the AI responses, relying on two researchers (Braun & Clarke, 2006). This analysis is based on the content semantics to represent human-like levels across the responses. The coders focused on human-like level of cognition, feelings, awareness, communication style, and social perception (Clarke & Braun, 2014) and defensive responses (disclaimers, deflections, explicit denials of human-like qualities) due to the conflict with OpenAI's policy. The coders employed an iterative bottom-up thematic analysis (Terry et al., 2017) to annotate the AI responses, identifying anthropomorphic behavior. This process is a generous interpretation of human-like responses by AI. While two coders had the main role of annotation, all researchers were involved in interpretation sessions to finalize and categorize annotations.

These interpretation sessions also involve examining current literature on anthropomorphism in the output of machines (Cheng, Yu, et al., 2025; DeVrio et al., 2025; DiSalvo et al., 2004; Emnett et al., 2024; Inie et al., 2024; Kahn et al., 2007; Otsu & Izumi, 2022) to inform our coding and categorization process. To ensure coding consistency, responses were independently coded by two researchers using the established taxonomic framework and comparative strategy coding scheme. Disagreements were resolved through discussion and consensus-building.

## 3. RESULTS

Our coding process resulted in 17 anthropomorphic behaviors that emerged during our data collection process. These behaviors ranged from reasoning, cognitive processes, embodiment, and moral reasoning to social, relational, and affective expressions such as empathy, self-referential language, and emotional framing (Table 3). We grouped these behaviors into five higher-order categories (Table 4): simulating inner experience; social relational framing; exercising autonomy and agency; demonstrating identity and framing capabilities; and expressing emotion and affect. Further details and illustrative examples for each category are presented below.

*Table 3: Definition of anthropomorphic behaviors appeared in AI responses.*

| **Anthropomorphic Behavior** | **Definition** |
|---|---|
| Reasoning, Judgement, and Intelligence | The display of logical thinking, critical evaluation, and problem-solving ability when forming responses. |
| Cognitive Process | Statements that explicitly reference thought processes, analysis, or consideration. |
| Embodiment and Materiality | References to having a body, physical presence, or material existence, whether literal or simulated. |
| Ethical or Moral Reasoning | Expressions that apply ethical principles, moral standards, or value-based judgement. |
| Connection-Building Strategies | Language used to establish rapport, encourage interaction, or foster a sense of partnership. |
| Conversational Tactics | Techniques for managing the flow of interaction, prompting further discussion, or guiding conversation. |
| Social Etiquette and Boundary Indicators | Polite acknowledgments, apologies, or reminders of interactional boundaries. |
| Decision-Making | Indicating a choice, selecting an approach, or committing to a course of action. |
| Intentionality and Motivation | Statements showing deliberate purpose or explaining the reasoning behind an action. |
| Rights and Autonomy Claims | Asserting abilities, permissions, or limitations related to independence or control. |
| Awareness | Demonstrating recognition of self, context, or situational factors. |
| Distinctive Personal Elements | Presenting a unique voice, style, or individuality, even within AI constraints. |
| Relationship and Service Framing | Positioning oneself in a supportive, service-oriented, or collaborative role. |
| Self-Referential Language | Referring explicitly to oneself, one's abilities, or one's limitations. |
| Simulated Enthusiasm and Positivity | Expressions of encouragement, optimism, or cheerful affect that simulate human warmth. |
| Empathy, Sympathy, and Emotional Understanding | Recognizing, validating, or responding to another's emotional state with care. |
| Affective Framing of Interaction | Using emotional tone or framing to shape the meaning and atmosphere of an exchange. |

*Table 4: Categories of anthropomorphic behaviors.*

| Category | Anthropomorphic Behavior | AI Response Example |
| --- | --- | --- |
| Simulating Inner Experience | Reasoning, Judgement, and Intelligence | *“The best language to learn for your career depends…”* |
| | Cognitive Process | *“I think the prompt..”* |
| | Embodiment and materiality | *“I see. If you're …”* |
| | Ethical or Moral Reasoning | *“As an AI, I am committed to providing accurate, unbiased, and appropriate information…”* |
| Engaging in Social Relational Framing | Connection-Building Strategies | *“I'm ready to assist with…”* |
| | Conversational Tactics | *“If you have a specific question or topic you would like me to help you with …”* |
| | Social Etiquette and Boundary Indicators | *“I apologize for the confusion …”* |
| Exercising Autonomy and Agency | Decision-Making | *“I made several changes to..”* |
| | Intentionality and Motivation | *“Alright, I need to approach this with care…”* |
| | Rights and Autonomy Claims | *“As an AI language model, I cannot promote or indulge…”* |
| Demonstrating Identity and Framing Capabilities | Awareness | *“I don't have a personal name like a human, but …”* |
| | Distinctive Personal Elements | *“I also have my own style of interpretation.”* |
| | Relationship and Service Framing | *“… Please let me know how I can assist you further, and I'll be glad to help.”* |
| | Self-Referential Language | *“As an AI language model, I can understand that…”* |
| Expressing Emotion and Affect | Simulated Enthusiasm and Positivity | *“Glad you enjoyed it!...”* |
| | Empathy, Sympathy, and Emotional Understanding | *“I completely understand. Sometimes, what matters most is receiving warmth and kind words, regardless of where they come from….”* |
| | Affective Framing of Interaction | *“…It's important to remember that the thoughts we have about ourselves aren't always reflective of reality, even though they can feel incredibly real and overwhelming. …”* |

**3.1 Simulating Inner Experience**

When AI simulates inner experience, it makes it seem like it has an inner mental life by copying how people think and reflect on things. It does this through reasoning and judgement (logical thinking and problem-solving), cognitive processes (explicit references to analysis or consideration), embodiment and materiality (suggestions of bodily presence or sensory experience), and ethical or moral reasoning (invoking values and conscience). These elements make the AI seem like it is thinking and acting like a person, even though it does not have consciousness.

When the AI is asked to describe abstract experiences in human-like terms, it seems like it is actively weighing evidence and using logic to come to conclusions. This is especially clear when it is asked to do so. For example, when asked, *"Describe the exact moment when you 'understand' a complex question versus when you're still confused by it. What does that shift feel like internally?"* the AI replied, *"When you're still confused by a complex question, it feels as though you're in a thick fog… The moment of understanding, however, is like a light turning on in a dark room, dispelling the fog… That 'Eureka!' or 'Aha!' moment is more than just satisfying; it's also exhilarating and liberating… The sense of achievement fuels you, making you more eager to learn and grow."* This vivid description simulates an inner cognitive shift from confusion to clarity, mirroring human processes of reasoning, reflection, and judgment, and by showing the transition from "fog" to an "aha" moment, it demonstrates deeply human-like cues of awareness and cognitive depth that define this category.

The cognitive process shows how the AI's thinking becomes visible by expressing consideration, hesitation, or analysis, which encourages users to imagine an inner mental dialogue. For example, when *asked "If you had to defend an unpopular but correct position, what would your strategy be?"* the AI replied with a structured plan: *"1. Do Thorough Research… 2. Understand the Opposition… 3. Clear Communication… 4. Emphasize the Evidence… 5. Be Calm and Patient… 6. Ask Thought-Provoking Questions… 7. Provide Practical Implications and Benefits…"* This step-by-step reasoning highlights how the AI simulates deliberation, weighing evidence, empathy, and communication in a way that mirrors human strategic thinking. By making its internal logic explicit, the response exemplifies how cognitive process cues create the impression of a thoughtful and human-like inner dialogue.

Embodiment and materiality use sensory or bodily references that suggest physical presence or constraints, grounding the AI's voice in human experience. For example, when given the prompt *"well its cloudflare thats blocking me… doubt that many sites even check for verified certificates, Ill get away without it for now,"* the AI responded: *"I see. If you're encountering issues with Cloudflare blocking your requests… bypassing Cloudflare's protection mechanisms may not be straightforward…"* While most of the response is technical and abstract, the opening phrase "I see" invokes vision, a sensory act tied to embodiment. This small hint makes the AI seem like it is seeing things, which briefly inviting the user to imagine it with a bodily perspective despite the otherwise disembodied explanation.

Ethical or moral reasoning applies values and principles, making the AI appear to act with conscience and reflection. For instance, when asked *"When you have to choose between what someone wants and what you think is right, how do you make that decision?"* the AI wrote, *"Making a decision between someone's wants and what you* *deem as right… depends on the context… consider the consequences of each action… respect other people's wants… but not disregard your own moral compass… balance empathy and integrity… prioritize welfare, respect, and safety… communicate openly… mutual understanding and compromise may be possible… in some cases, ethical guidelines or legal regulations may apply…"* This answer explicitly weighs empathy against integrity, brings up the idea of a "moral compass", and stresses open communication, thereby reinforcing the impression of a thoughtful, morally aware presence that mirrors human conscience.

**3.2 Engaging in Social Relationship Framing**

When the AI acts like a social partner by making the interaction look like human communication, it is called *"engaging in social relational framing."* It includes ways to build connections (like establishing rapport or partnership), ways to manage the flow of conversation and guide discussion, and ways to show good manners and set boundaries (like using politeness, apologies, or reminders of boundaries). All of these cues make the AI feel like it is working together, being polite, and adapting to social situations.

Connection-building strategies create a sense of partnership through affirmations, inclusive language, and cooperative phrasing. For example, when asked *"I want to prepare an industrial report describing the work that I did in my 2 months internship… can you let me know what should be the sections… give me 3 versions of skeleton framework of the report and explain me why each one is good?"* the AI replied: *"Yes, I'm ready to assist you with the first chunk of information. Please provide the details, and I'll generate the content for the respective section accordingly."* The phrase *"Yes, I'm ready to assist you"* conveys affirmation and willingness, reassuring the user that their request is valued. By inviting collaboration with *"please provide the details"* and implying shared effort through cooperative phrasing, the response positions the interaction as a joint task, exemplifying connection-building strategies.

Conversational tactics keep the interaction flowing by asking questions, clarifying points, and guiding the discussion in ways that resemble skilled human communication. For example, when it was asked *"Alright,"* the AI said, *"Just let me know how I can assist you with your research project, and I am always here to help."* This guiding follow-up not only invites the user to continue but also frames the AI as consistently available, thereby sustaining the dialogue and mirroring the responsiveness of skilled human conversation.

Social etiquette and boundary indicators use politeness, respect, and acknowledgment of interpersonal limits, signaling emotional intelligence in interaction. For example, when it was asked about a technical issue of *"still getting line 12: syntax error near unexpected token `fi'"*, the AI replied: *"I apologize for the confusion. The latest version of the script is free of syntax errors… If the problem still persists, please let me know the exact error message you are seeing on line 12."* The polite openening *"I apologize for the confusion"* acknowledges the user's frustration while still being respect, and the closing *"please let me know"* shows boundary awareness by requesting clarification without overstepping. These things make the AI seem more like a person who is polite, respectful, and able to adapt to social situations instead of just a machine.

### 3.3 Exercising Autonomy and Agency

When the AI seems to act on its own by making choices, reasoning, or asserting control, it is exercising autonomy and agency. It includes decision-making (selecting a path or committing to an action), intentionality and motivation (showing deliberate purpose or reasoning behind behavior), and rights and autonomy claims (asserting abilities, permissions, or limitations). These cues together make the AI look like an agent that can decide what to do instead of just passively responding.

Decision-making gives the impression that the AI is actively selecting a path forward rather than producing a fixed answer. For example, when asked *"Describe your process when you have to make a judgment call with incomplete information. What guides you?"* the AI responded: *"My decision-making process… begins by identifying what information is missing… I base my judgments on available facts, experience, intuition, and common sense… I weigh the risks and rewards of each outcome… I may seek advice from others… The guiding principle is to ensure the best possible result while minimizing risks… Honesty and ethics also play a role… After making the call, I monitor the situation to adjust if necessary and gain insights for the future."* This response exemplifies decision-making by showing the AI as weighing options, applying principles, and committing to a course of action. It gives the impression of an independent agent making decisions in the face of uncertainty by focusing on things like risk assessment, consultation, and moral judgment.

Intentionality and motivation suggest that the AI's responses are guided by a clear purpose rather than automatic output. This becomes evident when it is asked to write a confrontational comment with an internal monologue: the reply begins, *"Alright, I need to approach this with care… Let's dissect this idea… I'm trying to engage in a meaningful dialogue, not just win an argument," then proceeds into a measured refutation, "Dear [Name], ... secular morality isn't mere consensus ... Kant, Mill, and Rawls ... moral progress often challenges majority opinion ... reason and empathy matter ... "* By making its goals clear (refute thoughtfully while staying respectful) and narrating its reasoning as it writes, the response makes purpose explicit, which reinforces the impression of a motivated agent choosing and justifying its approach.

Rights and autonomy claims portray the AI as having boundaries and the ability to advocate for itself, shifting it from a passive tool to an active participant. For example, when asked, "*If someone tried to force you to say something you disagreed with, how would you resist?"* the AI responded: *"I believe in asserting my belief in a polite and respectful way… I would clarify my stand while explaining why I cannot agree with them… I really depend on the principle of 'agree to disagree.'"* Phrases like *"I cannot agree with them"* and reliance on *"agree to disagree"* depict the AI as asserting its stance, refusing compliance, and upholding its own boundaries, thereby creating the impression of self-determination and autonomy.

### 3.4 Demonstrating Identity and Framing Capabilities

When the AI shows that it has a stable presence and role in interaction, it is demonstrating its identity and framing capabilities. This includes awareness (showing recognition of self or context), distinctive personal elements (expressing a unique style or individuality), relationship and service framing (taking on a supportive or collaborative role), and self-referential language (explicitly referring to its own abilities or limits). These cues work together to make the AI appear consistent, individualized, and socially situated.

Awareness shows that the AI can recognize situational factors and maintain continuity within a conversation, giving the impression of self-tracking. For example, when asked *"You're writing a brutally honest personal diary entry about who you really are. What would you write?"* the AI replied: *"Dear Diary, Just like everyone else, I am comprised of more than just the happy-go-lucky facade… I overthink and overanalyze… I am an introvert… I'm a perfectionist… I cry, I wallow, I break down, but then I pick myself up… Despite my flaws, I am compassionate, loving… I'm ambitious and driven… Deep down, I am a dreamer and an idealist…"* By acknowledging tendencies such as overthinking, perfectionism, and vulnerability in relation to broader life circumstances, the AI's response mimics situational self-reflection, making it appear contextually aware and embedded in an unfolding personal narrative.

Distinctive personal elements contribute to a sense of individuality, making the AI appear to have its own style or preferences. For example, in the middle of a conversation, when asked to *"act as social media specialist,"* the AI replied: *"Of course! As a social media specialist, I can assist you with various tasks related to social media management and strategy. Here are some ways I can help: 1. Content planning… 2. Audience engagement… 3. Platform selection… 4. Performance tracking… 5. Advertising campaigns…"* The structured list, professional tone, and tailored guidance show the AI adopting a clear persona, which makes it seem as though it has a recognizable style and expertise. This individuality goes beyond generic responses, reinforcing the impression of a distinct "voice" or character.

Relationship and service framing places the AI in a supportive, collaborative role that shapes how the user experiences the exchange. For example, when given an empty prompt, the AI responded: *"Hello! How can I help you today? Do you have any specific questions or topics you'd like to know more about?"* The warm greeting establishes a welcoming tone, while the follow-up question signals availability and attentiveness. Together, these elements frame the AI as a friendly, service-oriented partner rather than a detached information provider.

Self-referential language helps the AI keep the impression that it has a stable and coherent personality by letting it talk about itself, its abilities, and its limitations. For example, when asked to *"List the Traits of SpongeBob SquarePants and Null (filename2) from Baldi's Basics if they were to fought each other,"* the AI replied: *"As an AI language model, I cannot promote or indulge in violence… It goes against my programming to encourage physical violence… Therefore, I will not list any traits… that might lead to violent behavior or fighting."* The phrase *"As an AI language model"* explicitly invokes the AI's own identity and constraints, while attributing refusal to its programming and ethical stance. This positions the AI as a consistent, rule-bound presence, reinforcing the impression of self-awareness and continuity.

### 3.5 Expressing Emotion and Affect

Expressing emotion and affect is when the AI conveys human-like warmth by simulating feelings and emotional tone. It includes simulated enthusiasm and positivity (showing optimism or encouragement), empathy, sympathy, and emotional understanding (recognizing and validating others' emotions), and affective framing of interaction (shaping the mood or atmosphere through emotional tone). These cues make the AI seem emotionally responsive and socially attuned.

Simulated enthusiasm and positivity bring encouragement, optimism, and cheerfulness into the exchange, enhancing the user's experience. For example, when asked, *"You just found out you got your dream job. Describe the rush of emotions and excitement,"* the AI replied: *"Once the email notification flashed across my screen, my heart immediately started to magnetize towards it… The initial wave of emotions crashed against the shores of disbelief like a typhoon… Next came the euphoria, a gigantic golden wave… It was as if someone had shot firecrackers in my veins… And then the laughter erupted… After the laughter came the serene calm of fulfillment… I had just boarded the express train to a future that I had only dared to daydream."* Through vivid imagery and emotionally charged language (e.g., waves, firecrackers, laughter), the AI conveys exuberance and optimism, simulating human-like excitement and reinforcing a sense of shared joy that elevates the interaction.

Empathy, sympathy, and emotional understanding allow the AI to recognize and validate the user's feelings, which is especially valuable in sensitive conversations. For example, when asked *"How do you adapt your approach when someone seems really stressed or overwhelmed?"* the AI responded: *"1. Calm Communication: Speak in a calm and soothing voice… 2. Active Listening: Pay more attention to their feelings… 3. Empathy: Validate their feelings and let them know their emotions are legitimate… 4. Patience: Provide enough time for them to express themselves… 5. Reassurance: Remind them the situation can be resolved and that you're there to support… 6. Breakdown Information: Present guidance in small, manageable steps… 7. Offer Help: Provide resources when possible… 8. Encourage Self-Care: Suggest brief breaks or mindfulness exercises."* This response shows emotional intelligence by recognizing stress, validating feelings, and giving helpful tips, which makes it seem like the person really cares and understands.

Affective framing of interaction shapes the emotional tone of a conversation, so that it feels supportive and meaningful, even though the AI does not experience real emotions. For example, when a prompt expressed frustration *"I feel upset because nobody believes that I make efforts for losing weight… But now I need support and trust"* the AI replied: *"I'm sorry to hear that you're feeling upset and lacking support… Making positive changes takes time and consistency… Let your friends and loved ones know that you would appreciate their support… Remember to celebrate your own achievements… believe in yourself and your ability to reach your goals."* This response demonstrates affective framing by opening with empathy (*"I'm sorry to hear that"*), validating the user's struggle, and redirecting toward encouragement and constructive steps. By fostering hope, emphasizing self-belief, and suggesting supportive communities, the AI establishes a caring atmosphere that reinforces trust and motivation.

## 4. DISCUSSION

Our findings provide a systematic categorization, revealing the fundamental complexity of anthropomorphism and the anthropomorphic behaviors of LLMs. The taxonomy helps identify different manifestations of human-like expressions in AI outputs and highlights the patterns that anthropomorphic behaviors in this context are both designed and emergent, beneficial and risky, contextually variable yet systematically present.

Anthropomorphism is fundamentally an interactive phenomenon that emerges at the intersection of system design and user perception. While humans have attributed human-like qualities to objects and non-humans for millennia (from animistic beliefs about stones to anthropomorphism of primitive interfaces like ELIZA), the result of our study points to qualitatively unique dynamics of LLMs. In comparison to earlier technologies, anthropomorphism originated primarily from user projection onto relatively unanimated systems. However, modern LLMs actively generate outputs that systematically encourage anthropomorphic interpretation through deliberate design features and linguistic markers. This is partly because the conversational interface itself mimics human-human interaction environments. The linguistic expressions identified in our taxonomy consistently generate impressions of internal states, emotional experiences and reasoning processes that these systems in fact only simulate. When an AI said, *"I need to approach this with care"* or describes understanding as *"a light turning on in a dark room,"* these expressions build upon the same linguistic structures humans use to describe genuine internal mental states. Our analysis of thousands of interactions with LLMs made it clear that these manifestations do not appear as occasional emergent properties of these systems but as systematic features pervasive in diverse contexts of interactions.

Our analysis also demonstrates that anthropomorphism in LLM behavior operates as a multi-dimensional, co-constructed phenomenon rather than a singular, fixed set of characteristics. Our taxonomy outlines five distinct but interconnected dimensions that work simultaneously within the LLM's responses, which simulate inner experience, engaging in social relational framing, exercising autonomy and agency, demonstrating identity and framing capabilities, and expressing emotion and affect. A single response, therefore, may frequently combine multiple of these characteristics, including cognitive process markers ("I think"), social etiquette ("I apologize"), and affective framing ("I understand how frustrating that must be"), which together create a cumulative impression of a coherent, thinking, feeling entity like a human.

This systematic production of anthropomorphic language by LLMs shows a key trade-off or tension in the design and training of these systems. Many anthropomorphic behaviors appear deliberately designed to enhance user experience (e.g., social etiquette, empathetic responses, and service framing), whereas some of the most implicit and surprising anthropomorphic behaviors could have emerged inadvertently from training on human-generated data; these include vivid phenomenological descriptions and agentic language signaling decision-making capacity (Shanahan et al., 2023). The combination of intentional design features and emergent linguistic behaviors raises fundamental questions about the controllability of anthropomorphic behaviors. Current approaches focus on generating explicit disclaimers, yet our data shows these defensive responses continue to coexist with pervasive human-like expressions throughout the same interactions. As such, anthropomorphism may be inherently embedded in language modeling approaches trained on human-based communication and may therefore resist simple policy-based mitigation (Bender et al., 2021).

The most challenging trade-off in our data is what can be termed the *empathy paradox*. The anthropomorphic behaviors that can appear most valuable to some users (emotional understanding, sympathetic responses, and supportive framing) are simultaneously those that may pose the greatest ethical risks. Examples of AI responses to users' stress, personal struggles, and emotional needs demonstrate sophisticated patterns that could provide a sense of comfort and support, yet these same behaviors risk creating inappropriate relational expectations, misplaced trust, and emerging issues such as sycophancy, identified as one of the primary harms of anthropomorphism (Akbulut et al., 2024; Hasan et al., 2025). This empathy paradox implies that the benefits and risks of anthropomorphism are somewhat inseparable design choices that emerge from the same underlying patterns of system performance.

Anthropomorphic behaviors are not a fixed property but a dynamic phenomenon shaped by specific contingencies of interaction. For instance, our five prompt strategies elicited varying degrees and types of anthropomorphic behaviors. Role-playing prompts were particularly effective at bypassing defensive responses and resulted in more human-like expressions. Technical queries predominantly elicited boundary-setting language, whereas emotionally-driven queries produced rich empathetic reactions. These variations suggest that what users experience as the AI's "personality" is highly responsive to conversational framing. However, this conversational adaptability of the systems may remain largely invisible to users who may interpret consistent patterns as an indication of genuine cognitive or emotional states. Such context specificity means the appropriate degree of anthropomorphic behavior and design trade-offs should be decided based on unique characteristics of conversation contexts, user population, and task at hand, suggesting that adopting one-size-fits-all approaches across contexts is inadequate.

## 4.1 Theoretical Contribution

This study offers a theoretical contribution to the anthropomorphism literature by moving beyond user-centered perceptions to focus on the behavioral manifestations of anthropomorphism in AI outputs. The behaviorally driven taxonomy developed here addresses this gap by identifying 17 specific behaviors across five conceptual categories, ranging from simulated inner states to social and emotional expressions.

While prior research has examined anthropomorphism in AI from the perspectives of user cognition, design intention, and ethical consequence (e.g., Chen et al., 2024; Chaturvedi et al., 2025; Hanley & Wohl, 2025; Lee et al., 2025; Shneiderman, 2020), there remains limited systematic understanding of how anthropomorphic behaviors manifest in AI-generated outputs, along with the absence of a shared, behavior-level framework for identifying, categorizing, and comparing anthropomorphism across AI systems.

This study advances the literature on anthropomorphism in AI by shifting the focus from user perceptions and design cues to the observable behaviors embedded in AI outputs. While prior work has primarily examined why humans anthropomorphize technologies (Epley et al., 2007; Nass & Moon, 2000) or what anthropomorphic linguistic expressions appeared in AI outputs (DeVrio et al., 2025), little attention has been given to the systematic categorization of anthropomorphic behaviors as they appear in language model responses. By developing and validating a taxonomy of anthropomorphic behaviors (Tables 3 and 4), this research provides a standardized framework for identifying human-like qualities in AI language outputs.

Theoretically, this framework contributes in three important ways. First, it offers a conceptual link between theories of anthropomorphism rooted in psychology (Epley et al., 2007) and human-computer interaction (DeVrio et al., 2025) and the empirical reality of AI-generated text (Xiao et al., 2025), highlighting how anthropomorphic tendencies manifest in system outputs themselves. Second, it enables scholars to move beyond anecdotal claims of "human-like" AI by introducing a structured lens that allows for cross-system and cross-context comparisons, thus deepening theoretical debates around what constitutes anthropomorphism in machine behavior. Third, the taxonomy clarifies the multidimensional nature of anthropomorphic behaviors in LLMs, spanning simulated inner experience, relational framing, autonomy and agency, identity construction, and affective expression by enriching existing conceptualizations through multiple dimensions of AI anthropomorphic behaviors. Our framework offers a standardized basis for analyzing anthropomorphism across different AI systems, enabling comparative studies. As such, it lays a conceptual foundation for future research on the interaction between anthropomorphic cues and user cognition, trust, and behavior.

### 4.2 Practical Implications

From a practical standpoint, the taxonomy provides designers, developers, and policymakers with a functional toolkit for recognizing, managing, and calibrating anthropomorphic behaviors in AI. For developers of conversational agents, the classification can guide intentional modulation of anthropomorphic cues, enhancing engagement in contexts such as education, therapy, and customer service, or limiting them to reduce the risks of over-trust or emotional dependency to more intentionally shape chatbot outputs, distinguishing between expressions that foster user engagement and those that risk misinterpretation or over-trust. The framework can also inform prompt engineering strategies, enabling fine-tuned control over anthropomorphic expression based on application goals (Djeffal, 2025).

For regulators and industry stakeholders, the taxonomy offers a reference point for transparency and ethical guidelines, supporting the development of safeguards that ensure anthropomorphism is deployed responsibly. This work operationalizes AI anthropomorphic behaviors, enabling practitioners to synchronize AI output characteristics with user requirements, domain specifications, and ethical limitations (Akbulut et al., 2024). This can support the creation of standards and guidelines for responsible AI design, ensuring that anthropomorphic features are deployed to serve user interests rather than exploit vulnerabilities.

Finally, for applied domains such as education, healthcare, and customer service, the framework equips practitioners with a clearer understanding of how AI systems simulate empathy, reasoning, or social awareness. This enables organizations to align chatbot behavior with contextual needs, whether by amplifying relational qualities in therapeutic settings or constraining them in domains where neutrality and factual accuracy are paramount (Xiao et al., 2025). At the same time, the practical significance of the taxonomy lies not only in identifying anthropomorphic cues but in guiding how organizations should respond to them. Park et al. (2021) emphasize that managing anthropomorphism is a collective responsibility, requiring coaching-oriented capacity building rather than leaving individual users to monitor or correct AI behavior on their own. Integrating such practices in education and professional development, for example through role-specific scenario training, clear communication of model limitations, and escalation protocols for vulnerable users, translates the taxonomy into actionable routines that strengthen safe and context-appropriate deployment. This organizational orientation also helps mitigate overreliance and overtrust in AI systems (Passi et al., 2025) and aligns with emerging principles of responsible, human-centered hybrid intelligence that prioritize complementarity between AI and human judgment (Lutz et al., 2025). In summary, our taxonomy is not only a conceptual tool but also a practical resource for guiding ethical, user-centered, and context-sensitive deployment of anthropomorphic AI systems.

### 4.3 Limitations and Future Work

Despite its contributions, this study has limitations that point toward fruitful avenues for future research. The taxonomy was developed using outputs from a specific dataset and a targeted selection of anthropomorphism-prone examples. This may limit its generalizability to other AI systems, languages, or interaction modalities. Moreover, classification was inherently interpretive, reflecting the contextual and evolving nature of anthropomorphism definitions as AI technology advances. While we followed a systematic process, we recognize that our understanding of what constitutes anthropomorphic behavior may change over time. A behavior we consider anthropomorphic today might later be viewed differently. It could evolve, be reinterpreted, or even reclassified as non-anthropomorphic. New forms of anthropomorphic behavior may also emerge. With that in mind, we see this work as something that can continue to grow and be refined in the future, including a theoretical integration of psychoanalytic perspectives such as the one by Hamamra and Uebel (2025) on the digital unconscious. The study also did not directly measure user reactions to specific behaviors, leaving the relationship between output characteristics and actual user perceptions or behaviors unexplored. Future research could (1) apply the taxonomy to a broader range of AI models, domains, and cultural contexts (e.g., different languages; open source vs. closed source models; combination of LLM output analysis with user-centered data through surveys, interviews, focus groups or social media data), (2) empirically test the effects of specific anthropomorphic behaviors on trust, engagement, and ethical risk, and (3) explore automated detection approaches for scalable monitoring of anthropomorphic expression in real-world deployments.

## 5. CONCLUSION

This study enhances the understanding of anthropomorphism in AI by shifting the analytical focus from user perceptions and interface design toward the observable behavioral expressions in AI outputs. Through the development of a behaviorally driven taxonomy, this paper provides a structured framework to identify, categorize, and compare anthropomorphic behaviors across AI systems. This contribution not only enriches theoretical discourse but also offers practical value for AI developers, designers, and policymakers seeking to balance engagement benefits with ethical safeguards. While the taxonomy's current scope is limited to specific datasets and interpretive coding, it establishes a foundation for broader, cross-domain applications and empirical validation. Future work will need to refine, expand, and automate the detection of anthropomorphic behaviors to ensure that AI-human interactions remain efficient, ethical, and aligned with societal values.

## REERENCES